\documentclass[twocolumn,floats,aps,amsmath]{revtex4}
\usepackage{rotating}
\usepackage{graphicx}% Include figure files
\usepackage{dcolumn}% Align table columns on decimal point
\usepackage{bm}% bold math
\usepackage{amssymb} %maths
\usepackage{amsmath} %maths
\usepackage[utf8]{inputenc} %useful to type directly accentuated characters

\begin{document}

\preprint{APS/123-QED}

\title{Non-linear effect of uniaxial pressure on superconductivity in
CeCoIn$_5$} 
\author{S. D. Johnson}
\affiliation{Physics Department, University of California Davis, Davis, CA
95616}
\author{R. J. Zieve}
\affiliation{Physics Department, University of California Davis, Davis, CA
95616}

\author{J. C. Cooley}
\affiliation{
Los Alamos National Laboratory, Los Alamos, NM 87545}
\date{\today}

\newcommand{\5}{$_5$ }
\newcommand{\tc}{$T_{c}$ }
\newcommand{\tcp}{$T_{c}(P)$ }
\newcommand{\tcpa}{$T_{c}(P_{app})$ }
\newcommand{\fp}{$f(p)$ }
\newcommand{\dtc}{$\Delta T_{c}$ }
\newcommand{\ca}{\emph{c/a} }
\newcommand{\dtcpa}{$\Delta T_{c}(P_{app})$ }

\begin{abstract}
We study single-crystal CeCoIn$_5$ with uniaxial
pressure up to 3.97 kbar applied along the \emph{c}-axis. We find a
non-linear dependence of the superconducting transition temperature
\tc on pressure, with a maximum close to 2 kbar.  The transition also
broadens significantly as pressure increases.  
We discuss the temperature dependence in terms of the general
trend that \tc decreases in anisotropic heavy-fermion compounds
as they move towards three-dimensional behavior.
\end{abstract}


\maketitle

\section{Introduction} 

The past three decades have seen the discovery of several new groups
of superconductors that do not conform to the previous 
understanding of superconductivity: the cuprates \cite{bm}, heavy fermions
\cite{CeCu2Si2}, organics \cite{bechgaard}, and the very recent pnictide
superconductors \cite{pnictide}.  Although behavior varies substantially
among these materials, in many cases even for closely related compounds,
similarities include deviations from Fermi liquid behavior \cite{varma,
moser, Stewart01, wu} and the appearance of superconductivity near the
quantum critical point where a magnetic transition is suppressed to zero
temperature \cite{mathur, kotegawa, taillefer}.  From these observations a
general picture is emerging of superconductivity with Cooper pairs bound
by magnetic fluctuations.  The magnetic interaction favors the $d$-wave
pairing symmetry which has been established in the cuprates \cite{tsuei}
and indicated elsewhere \cite{tou, stockert, kasahara, ichimura}.  Low
dimensionality also facilitates superconductivity \cite{monthoux2001,
monthoux2002}, since the pairing interaction falls off less quickly in
lower dimensions.

The 115 superconductors, CeMIn$_5$ (M = Co \cite{petrovicco},
Rh \cite{hegger}, Ir \cite{petrovicir}) and the isostructural PuMGa$_5$
(M = Co \cite{sarrao}, Rh \cite{wastin}) family, are useful materials for
testing effects of dimensionality.  Both families are heavy-fermion
superconductors that also exhibit antiferromagnetism, sometimes in
concert with superconductivity.  They are clean, relatively easy to grow,
and close to a quantum critical point at ambient pressure.  The crystal
structure is tetragonal, with alternating layers of CeIn$_3$ and MIn$_2$,
resulting in anisotropic superconductivity.  For the 115 materials,
including several alloys where $M$ is a mixture of two elements,
the superconducting transition temperature $T_c$ increases linearly
with $c/a$ within a family \cite{pagliuso}, with the same logarithmic
derivative $\frac{d(\ln T_c)}{d(c/a)}$ for both the Ce and Pu families
\cite{thompsoniso}.  In several cases uniaxial pressure adheres to
this trend.  By pushing the planes together, $c$-axis pressure should
decrease $T_c$.  On the other hand, $a$-axis pressure increases the
plane separation and should increase $T_c$.  For CeIrIn$_5$,
uniaxial pressure has exactly these effects, with measurements made
both directly \cite{dix} and by extracting the zero-pressure $dT_c/dP$
from thermal expansion data \cite{oeschler}.  However, for CeCoIn$_5$
thermal expansion measurements yield positive $\frac{dT_c}{dP}|_{P=0}$
for $c$-axis pressure as well as $a$-axis pressure \cite{oeschler}.
Here we apply uniaxial pressure along the $c$-axis of CeCoIn$_5$
to investigate this apparent exception to the pattern that higher
dimensionality corresponds to lower $T_c$.

\section{Materials \& Methods}

Single crystal samples were grown in aluminum crucibles containing
stoichiometric amounts of Ce and Co with an excess of In.  The crucibles
were sealed in quartz tubes, heated to 1150$^\circ$C, and slow cooled to 
450$^\circ$C.  Excess flux was removed by centrifuging.  CeCoIn$_5$
grows in thin platelets perpendicular to the crystal \emph{c}-axis.
To prepare samples for pressure measurements, we remove excess In
and polish the faces to be smooth and free from chips or defects.
The polishing also ensures that the faces used for pressure application
are parallel to each other.  We confirm the sample orientation by x-ray
diffraction, both before and after polishing.

We apply uniaxial pressure using a bellows setup activated with helium gas from
room temperature.  The pressure cell is permanently mounted on an Oxford
Instruments dilution refrigerator.  A piezo sensor monitors the force
in the pressure column (see \cite{dix} for pressure setup schematic).
With this setup we can reach a maximum pressure of about 10 kbar,
depending on the cross-sectional area of the sample, and we can change
pressure in controlled steps smaller than 0.1 kbar.

\begin{figure}[tb]
\begin{center}
\scalebox{.35}{\includegraphics{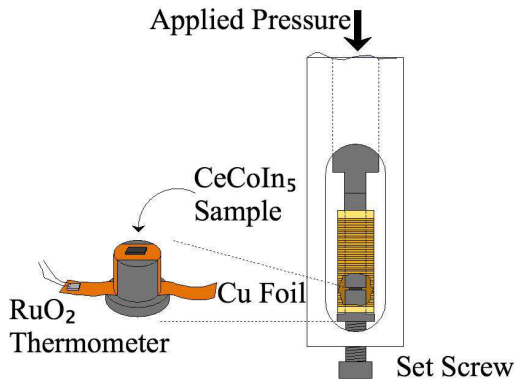}} 
\caption{A balanced ac-susceptibility coil is
placed around the pressure shafts.  The sample is between the two shafts
of the pressure column, centered in one of the pick-up coils. }
\label{setup}
\end{center}
\end{figure}

From the mass and thickness of the sample, we compute its cross-sectional
area, which we then use to calculate the pressure on the sample.
The data shown here are from a sample with mass 2.43 mg and area
1.54$\times 10^{-6}$ m$^2$.  Measurements of a second sample of mass
0.57 mg and area 3.63$\times 10^{-7}$m$^2$ agreed qualitatively, although
the smaller sample size reduced the quality of the data.

A screw at one end of the pressure column controls its overall length.  We
finger-tighten this screw while watching the pressure monitor, applying a
pressure of about 0.05 kbar to the sample at room temperature. This
ensures that the sample remains in place while we load the cryostat
into its dewar and cool from room temperature.  Thermal contraction
during cooling may alter the initial pressure; in previous work the initial
pressure appeared on the order of 0.3 kbar at low temperatures \cite{dix}.
The values for applied pressure used in this paper do not include any
offset for this initial pressure, but our initial transition temperature
measurements agree to better than 3 mK with measurements taken outside of
the pressure cell, suggesting a pressure offset of less than 0.2 kbar.
Once the cryostat is below 4 K we fill the bellows with liquid helium.
We then increase pressure in steps of about 0.5 kbar.  The maximum
pressure used here is 3.97 kbar.

In the present measurements, the effect of pressure is reversible.
The data shown here come from three pressure sweeps on the sample, the
first reaching a maximum pressure of 2.60 kbar, the second 3.97 kbar,
and the third 2.86 kbar.  After completing the measurements, we
confirmed by x-ray diffraction that the sample retained its original
crystal structure.

We detect the transition to the superconducting state with a balanced ac
susceptibility coil that accommodates the pressure shafts and the sample,
as shown in Figure \ref{setup}.  The outer primary coil is
0.9 inches long with inner diameter 0.35 inches.  It contains
approximately 600 turns of 0.006-inch diameter Cu wire and with our
usual settings generates a field of about 0.3 gauss parallel to the
pressure shafts.  The inner secondary coils are wound on a cylinder of
diameter of 0.16 inches and contain 200 turns each of 0.002-inch diameter
Cu wire.  The inner coils are separated by 0.325 inches, so that there
is minimal field interaction between coils.  We monitor the signal
from the inner coils using a Linear Research LR-700 resistance bridge.
We position the sample so that its entire volume is contained within
the bottom coil of the secondary.

To ensure that the sample is in thermal contact with the mixing chamber we
varnish Cu foil to the pressure shafts on either side of the sample, with
the other end of each foil varnished to the mixing chamber.  Even with the
Cu foil heat-sink, the large thermal mass of the pressure setup results in
a temperature lag between the sample and the mixing chamber during
temperature scans.  We monitor the sample temperature directly using
a RuO$_{2}$ thermometer mounted on the copper foil within 0.5 inches of
the sample.

\section{Results \& Discussion}

\begin{figure}[b]
\begin{center}
\scalebox{.47}{\includegraphics{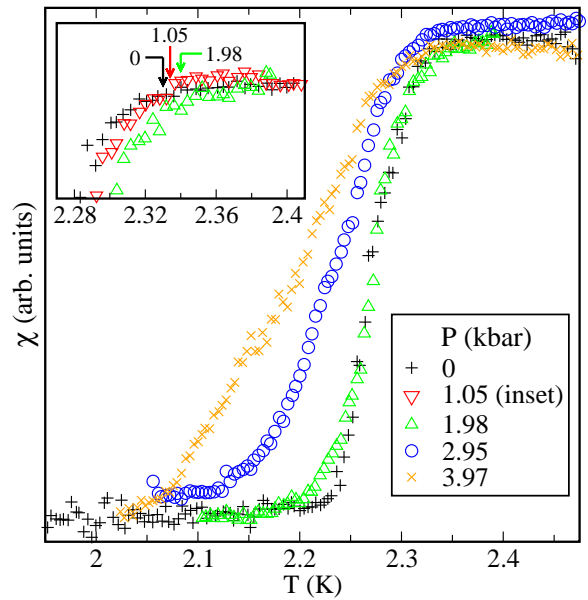}} 
\caption{Representative data showing the effect of $c$-axis
uniaxial pressure.  The transition width \dtc is strongly
pressure-dependent, while the onset temperature \tc is not.
The inset expands the region near \tc to reveal a slight increase in
\tc with pressure.
Arrows indicate the onset \tc for each curve.}
\label{data}
\end{center}
\end{figure}

\begin{figure}[thb]
\begin{center}
\scalebox{.43}{\includegraphics{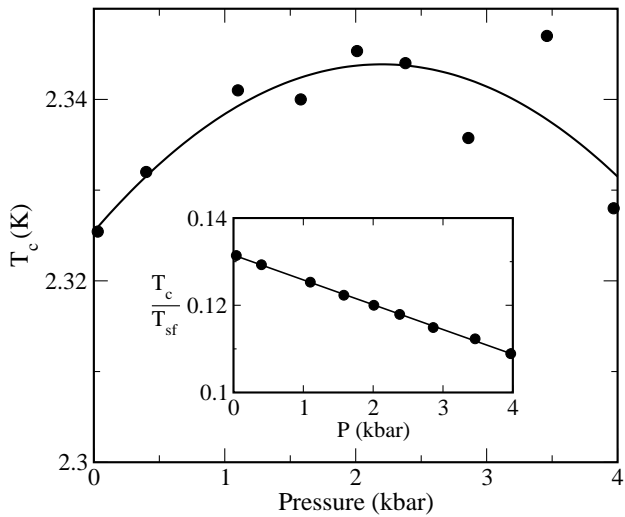}}
\caption{
Onset temperature for superconductivity as a function of pressure.  The
points are averages over measurements at similar applied pressures
$P_{app}$, and the curve is a quadratic fit.  Inset: $T_c/T_{sf}$ vs
pressure, with a linear fit shown.  As described in the text, $T_{sf}$
values are adapted from hydrostatic pressure measurements.
}
\label{tcp}
\end{center}
\end{figure}

Figure \ref{data} shows the superconducting transition for several
pressures.  The substantial broadening of the transition with pressure
ensures that the midpoint moves down in temperature with increasing
pressure.  However, as shown in the inset, the onset behaves differently,
initially moving to {\em higher} temperature as pressure increases.  Here we
identify the onset as the highest temperature at which the susceptibility $\chi$
deviates from its constant normal-state value $\chi_{n}$.  We plot the onset
temperatures \tc in Figure \ref{tcp}.  The parabola,
with a maximum near 2 kbar, is a least-squares fit.

\begin{figure}[tbh] 
\begin{center}
\scalebox{.43}{\includegraphics{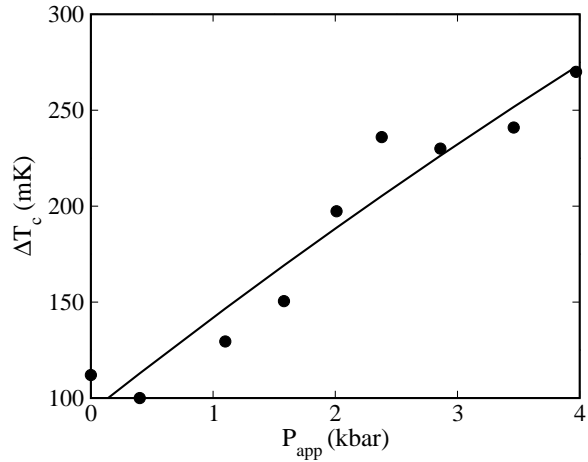}}
\caption{Superconducting transition width versus
applied pressure.  Solid line is the best linear fit to the data. } 
\label{dtc}
\end{center}
\end{figure}

The non-monotonic \tcp makes it particularly important to consider the meaning
of the transition width. The width $\Delta T_c$, defined as the difference
between onset of superconductivity and the leveling off of $\chi$ at the low end
of the transition, increases from 112 mK at zero pressure to 270 mK at 3.97
kbar, as shown in Figure \ref{dtc}. One possible source for the broadening is non-uniformity in the applied
pressure.  The pressure dependence of \tc would then lead to different
transition temperatures in different parts of the sample. Non-uniform pressure
could arise from a variety of effects, such as defects in the sample, variation
in its cross-sectional area, indium inclusions, surface irregularities,
a tilt in the pressure column, or an intrinsic inhomogeneity in the
pressure distribution within the sample.

We have estimated the pressure inhomogeneity needed to produce the observed
broadening, using various assumptions for the distribution of $c$-axis pressure
within the sample.  We find that to account for the entire increase in
transition width, the pressure would have to vary by more than a factor
of two across the sample.  This is a consequence of the scant change
ie the onset temperature of the transition.  Either the transition has
little pressure dependence, in which case variations of pressure would
not broaden it, or a portion of the sample remains at very low pressure
even when the nominal applied pressure is large.
While a non-constant sample cross-section or an angle of
the pressure spacers could cause inhomogeneity of a few percent, a 100\%
variation is far too extreme.  

The pressure could also vary in direction within the sample, creating
stress with an $a$-axis component.  However, as noted above, both thermal
expansion measurements \cite{oeschler} and the expected influence of
dimensionality suggest that $a$-axis pressure would increase $T_c$, which
would not explain the broadening of the transition to lower temperatures
that we observe.  We conclude that if the transition width truly signifies
pressure variation, it must indicate a broad distribution of pressure
within the sample from an intrinsic mechanism.

Interestingly, in resistivity measurements on CeCoIn$_5$ under hydrostatic
pressure the transition width has a minimum near
$P^*=16$ kbar, the pressure which maximizes $T_c$ \cite{sidorov}.
The transition width in specific heat exhibits a similar crossover
behavior, remaining nearly constant at low pressures but
increasing substantially once pressure exceeds 16 kbar \cite{knebel}.  The agreement
between the different types of measurement is evidence that the width
has an intrinsic component.  Here, our susceptibility measurements 
show a substantial increase in transition width from the lowest pressures.
If uniaxial $c$-axis pressure shifts CeCoIn$_5$ away from the $P^*$
reached with hydrostatic pressure, our observed broadening could plausibly
be a tuning effect related to the width changes observed with other techniques.

In the following discussion, we bypass concerns about the
origin of the transition width by focusing on the onset temperature.

The small initial slope of the \tc curve is consistent with a nearby
maximum.  Fits to our data suggest $\frac{dT_c}{dP}|_{P=0}=17$
mK/kbar. The value derived from thermal expansion measurements
\cite{oeschler} is even lower, 7.5 mK/kbar.  This pressure dependence
is substantially less than that of CeIrIn$_5$, where direct $c$-axis
pressure measurements give $\frac{dT_c}{dP}|_{P=0}=-66$ mK/kbar \cite{dix}
and thermal expansion suggests $\frac{dT_c}{dP}|_{P=0}=-89$ mK/kbar
\cite{oeschler}.  The pressure effect is also smaller than for $a$-axis
pressure in CeCoIn$_5$, where thermal expansion suggests that $T_c$
increases 29 mK/kbar \cite{oeschler}. One natural explanation is that
CeCoIn\5 at ambient pressure is near an extremum of $\frac{dT_c}{dP}$,
particularly for $c$-axis pressure.

Non-monotonic behavior is less common with uniaxial pressure than with its
hydrostatic counterpart.  Partly this is because the maximum pressure is
generally much smaller in the uniaxial case, due to sample breakage or to the
limits of the pressure apparatus.  In addition, a given hydrostatic
pressure may affect an isotropic sample in a similar way as three times
as much uniaxial pressure, a consequence of applying the pressure simultaneously
along all three perpendicular axes.  Together these considerations mean
that a typical uniaxial pressure measurement tunes a sample over a
narrow regime compared to standard hydrostatic techniques.

Hydrostatic pressure measurements on CeCoIn$_5$ \cite{sidorov} find a maximum
$T_c$ near 16 kbar.  Without anisotropic effects, one might expect
an equivalent uniaxial pressure to be 48 kbar, since hydrostatic
pressure involves stress applied along all three axes
simultaneously.  In fact, we find the maximum $T_c$ at a drastically lower
pressure near 2 kbar.  

The maximum in $T_c$ requires competing factors tending to raise or
lower $T_c$ with applied pressure.  The former is the hybridization
of neighboring atomic orbitals, which increases as pressure
reduces the atomic spacing.  Using a tight binding approximation
\cite{harrison,wills, kumar}, we estimate the fractional change in the
hybridization between the Ce $f$-electrons and the In $p$-electrons as 0.0665\%
per kbar of $c$-axis pressure.  An analogous calculation for CeIrIn\5 gives
0.0653\% change per kbar.  The similarity of these values implies that the main
difference between the materials lies elsewhere.

The other key factor is sample anisotropy, which decreases with $c$-axis
pressure.  This is consistent with our maximum $T_c$ occurring at a much lower
pressure than in hydrostatic pressure measurements, since hydrostatic pressure
has a more uniform effect on the sample.  Calculations also predict that lower
anisotropy should decrease superconducting transition temperatures
\cite{monthoux2002}.

The calculations track $T_c/T_{sf}$, where $T_{sf}$ is the spin-fluctuation
temperature that appears to set the energy scale in magnetically mediated
superconductors. In principle $T_{sf}$ is related to the normal-phase
susceptibility just above $T_c$.  However, that susceptibility is quite small
and changes only a few percent per kbar.  All told the changes in $\chi_n$ are
about five orders of magnitude smaller than the size of the superconducting
transition.  Our signal has comparable shifts from other factors, possibly
including small changes in sample shape and position with applied pressure. 
The lack of a measurement of $T_{sf}$ limits the comparison possible
with \cite{monthoux2002}. As a rough illustration of how the behavior
of $T_{sf}$ dominates that of $T_c$, we refer to data under hydrostatic
pressure \cite{nicklas}.  There $T_{sf}$, which is assumed proportional to
the temperature $T_M$ of the resistance maximum, increases about 6\% per
kbar.  We ignore anisotropy in the spin fluctuations and reduce the change
in $T_{sf}$ to 2\% per kbar to adjust between hydrostatic and uniaxial
pressure.  With our measured superconducting transition temperatures, we
then plot $T_c/T_{sf}$ as a function of applied $c$-axis pressure, shown
in the inset to Figure \ref{tcp}.  The pressure dependence of $T_{sf}$
dominates, changing the low-pressure maximum to a near-linear monotonic
decrease.  Although the actual dependence of $T_{sf}$ on $c$-axis pressure
may differ from this estimate, for any increase of roughly the same size
$T_{sf}$ mainly determines the behavior of $T_c/T_{sf}$.

That $T_{sf}$ increases with $c$-axis pressure is consistent with recent
experiments on CeIn$_3$/LaIn$_3$ heterostructures \cite{shishido}. As the
thickness of the CeIn$_3$ layers decreases, the effective mass increases,
an effect attributed to the changing dimensionality.  Our $c$-axis
pressure tends to increase dimensionality, which corresponds to a
decreasing effective mass and increasing $T_{sf}$.

\section{Conclusion}

We present ac-susceptibility measurements on a single crystal
sample of CeCoIn$_5$ under direct uniaxial pressure up to 3.97 kbar,
along the \emph{c}-axis.  We find a weak, non-linear dependence of \tc
on pressure.  After an initial increase to a maximum near 2 kbar,
\tc then decreases.  The decrease agrees qualitatively with the behavior
expected from decreasing the anisotropy parameter $c/a$.  We also find an
increase in transition width as pressure increases which is much larger
than would be expected from nonuniformity in pressure and may be connected
to our tuning through the superconducting phase.

\end{document}